\documentclass[aps,prl,twocolumn,superscriptaddress]{revtex4-2}
\usepackage{amsmath,amssymb,amsfonts}
\usepackage{bm, latexsym}
\usepackage{graphicx}
\usepackage{bbm,times}
\usepackage[normalem]{ulem}
\usepackage[usenames,dvipsnames]{color}
\usepackage[pdftex]{hyperref}
\usepackage{braket}
\usepackage{mathtools}

\begin{document}

\newcommand{\cred}[1]{\textcolor{black}{#1}}
\newcommand{\ccred}[1]{\textcolor{black}{#1}}

\title{Measurement-Induced Crossover of Quantum Jump Statistics in Postselection-Free Many-Body Dynamics}



\author{Kazuki Yamamoto}
\email{kazuki-yamamoto@omu.ac.jp}
\affiliation{Research Institute for Innovation and Co-Creation, Osaka Metropolitan University, Sakai, Osaka 599-8531, Japan}
\affiliation{Department of Physics, Osaka Metropolitan University, Sumiyoshi, Osaka 558-8585, Japan}
\affiliation{Nambu Yoichiro Institute of Theoretical and Experimental Physics (NITEP), Osaka Metropolitan University, Sumiyoshi, Osaka 558-8585, Japan}
\affiliation{Department of Physics, Institute of Science Tokyo, Meguro, Tokyo 152-8551, Japan}

\author{Ryusuke Hamazaki}
\affiliation{Nonequilibrium Quantum Statistical Mechanics RIKEN Hakubi Research Team, RIKEN Pioneering Research Institute (PRI), RIKEN iTHEMS, Wako, Saitama 351-0198, Japan}



\date{\today}

\begin{abstract}
We reveal a nontrivial crossover of subsystem fluctuations of quantum jumps in continuously monitored many-body systems, which have a trivial maximally mixed state as a steady-state density matrix. While the fluctuations exhibit the standard volume law $\propto L$ following Poissonian statistics for sufficiently weak measurement strength, anomalous yet universal scaling law $\propto L^\alpha \:(\alpha\sim 2.7$ up to $L=20)$ indicating super-Poissonian statistics appears for strong measurement strength. This drastically affects the precision of estimating the rate of quantum jumps: for strong (weak) measurement, the estimation uncertainty is enhanced (suppressed) as the system size increases. We demonstrate that the anomalous scaling of the subsystem fluctuation originates from an integrated many-body autocorrelation function and that the transient dynamics contributes to the scaling law rather than the Liouvillian gap. The measurement-induced crossover is accessed only from the postselection-free information obtained from the time and the position of quantum jumps and can be tested in ultracold atom experiments.
\end{abstract}

\maketitle

\textit{Introduction}--Dissipative quantum many-body phenomena induced by the coupling to environments offer a rich possibility to investigate physics unique to nonequilibrium situations \cite{Muller12, Harrington22, Fazio24}. One of the key ingredients is the measurement, which has been actively investigated in the context of open quantum systems \cite{Daley14, Ashida20}. \cred{Readout of measurement brings about nonequilibrium phenomena that cannot be seen in unconditional open quantum systems, such as measurement-induced entanglement transitions \cite{Fisher23rev, Fisher18, Smith19, Skinner19, Yaodong19, Noel22, Koh22, Google23, Cao19, Alberton21, Turkeshi21, Turkeshi22, Piccitto22, Tang20, Fuji20, Szyniszewski20, Lunt20, Jian21, Van21, Doggen22, Minato21, Muller21, Buchhold21, Yamamoto23, Marcin23, Mochizuki25, Matsubara25, Chakrabarti25}, unconventional non-Hermitian criticality \cite{Ashida16, Ashida17, Yamamoto22, Yamamoto23B}, and measurement-altered universality and boundary transitions \cite{Garratt23, Sun23, Tang24, Ashida24, Liu24}. While these works demonstrate the significance to study an interplay between measurements and many-body effects, the so-called postselection problem makes it difficult to experimentally observe measurement-induced quantum many-body phenomena, despite efforts to evade the problem \cite{Gullans20L, Yaodong23, Altman24, Ippoliti21, Grover21, Yamamoto23, Mog23, Passarelli24, Max24, Skinner25, Barratt22L1, Agrawal24, Ippoliti24, Akhtar24, Hansveer25}. Thus, it is of importance to investigate different type of measurement-induced physics that does not rely on postselections.}

\cred{Notably, in quantum simulations of many-body trajectory dynamics under continuous monitoring, we can access times and positions of quantum jumps without postselection \cite{Fazio24}. The statistics of such stochastic trajectories has been widely investigated both in classical \cite{Garrahan07, Garrahan17} and open quantum \cite{Raphael15, Garrahan18rev, Fresco24, Le24} systems, where various properties have been demonstrated such as dynamical phase transitions \cite{Garrahan07, Garrahan10} and thermodynamic uncertainty relations \cite{Barato15, Garrahan17, Carollo19, Horowitz20}.} For instance, full counting statistics of quantum jumps in open systems are useful to capture many-body physics \cite{Znidaric14, Hickey13, Lesanovski13, Carollo18A, Bao24} as well as few-body physics \cite{Landi24}. In particular, the variance of the number of jumps contains information that is not obtained from an average of physical quantities \cite{Landi24}; it is related with autocorrelations and relaxation rates \cite{Cenap12, Buca14, Znidaric14B, Znidaric14E}. For example, in driven dissipative many-body systems that break the detailed-balance conditions, diverging current fluctuations at the dissipative phase transition are studied \cite{Landi22, Matsumoto25}. Recently, it has been demonstrated that space-time correlations of measurement outcomes can reveal dynamical heterogeneity before thermalization in kinetically constrained many-body models \cite{Cech24}.

\begin{figure}[b]
\includegraphics[width=8.5cm]{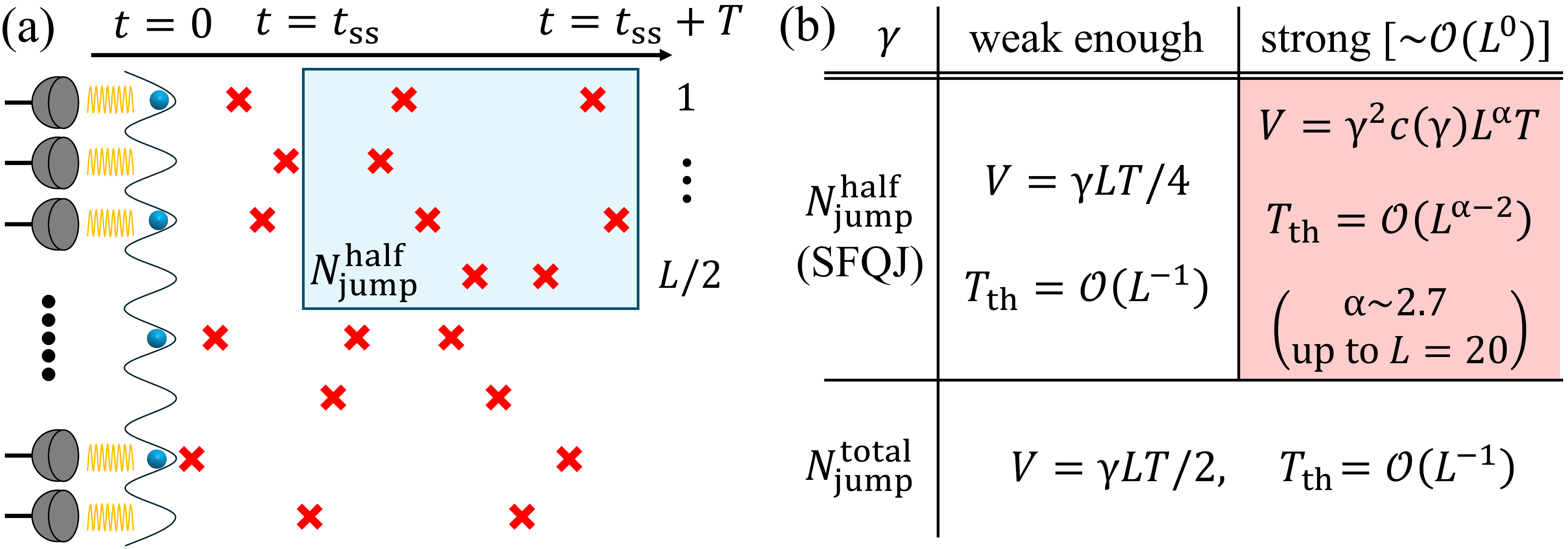}
\caption{(a) Schematic figure of our setup. We count the number of jumps (red cross marks) in a half chain $N_\mathrm{jump}^\mathrm{half}$ and take its variance along trajectory realizations. (b) Table of the main results for the variance $V$ and estimation uncertainty $T_\mathrm{th}$ concerning the number of jumps. Subsystem fluctuation of quantum jumps (SFQJ) for strong measurement exhibits anomalous yet universal super-Poissonian statistics, and the estimation uncertainty is enhanced as the system size is increased, in stark contrast to the case for the weak measurement or the total fluctuation of quantum jumps.}
\label{fig_schematic}
\end{figure}

\begin{figure*}[t]
\includegraphics[width=18cm]{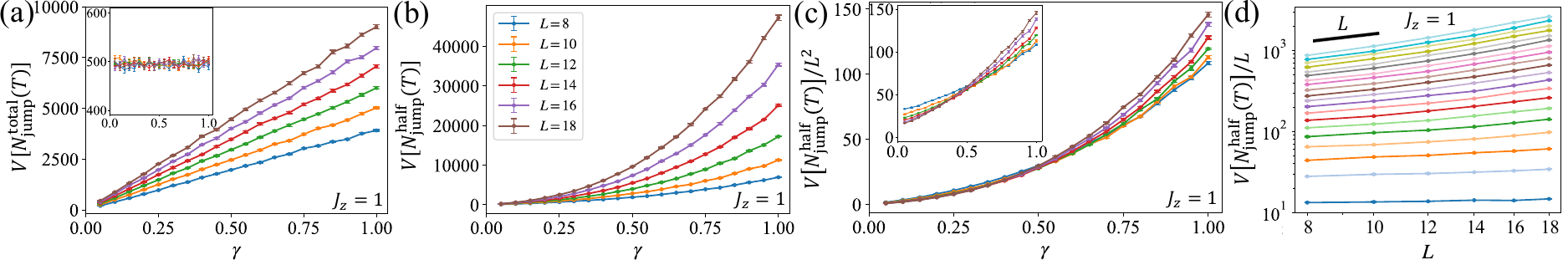}
\caption{Numerical results for the variance of quantum jumps for the Heisenberg model ($J_z=1$) for $832$ trajectories and $T=990$. (a) Variance of net quantum jumps in the whole system [Inset: $V[N_\mathrm{jump}^\mathrm{total}(T)]/(\gamma L)$], (b) SFQJ, (c) $V[N_\mathrm{jump}^\mathrm{half}(T)]/L^2$ [Inset: $V[N_\mathrm{jump}^\mathrm{half}(T)]/(\gamma L^2)$], and (d) $V[N_\mathrm{jump}^\mathrm{half}(T)]/L$. The SFQJ exhibits a measurement-induced crossover of the system-size scaling.  Data are plotted against $\gamma$ for system sizes $L=8, 10, \cdots, 18$ in (a)-(c) and against $L$ for measurement strengths $\gamma=0.05, 0.1, \cdots, 1$ from bottom to top in (d). We take the average over 20 time intervals for $t\in [200, 1190], [1190, 2180], \cdots, [19010, 2\times10^4]$ \cite{steadystate}.}
\label{fig_variance}
\end{figure*}

\cred{However, it is highly nontrivial whether the quantum jump statistics exhibit universal many-body phenomena in the steady state characterized by the (unconditional) density matrix that heats up to a thermal state, which naturally appears in quantum many-body systems under particle-number measurement \cite{Fuji20, Yamamoto23}. In particular, it is intriguing to investigate the impact of the measurement strength on fluctuations of quantum jumps even when the steady-state density matrix is independent of the measurement strength. This may lead to a novel measurement-induced phenomena distinct from, e.g., entanglement transitions.}

\cred{In this Letter, we introduce the subsystem fluctuation of quantum jumps (SFQJ) as a new indicator to investigate measurement effects of continuously monitored many-body systems and discover its crossover in the steady state.} In particular, we demonstrate that SFQJ obeys the Poissonian-type scaling $\propto L$ ($L$ is the system size) for sufficiently weak measurement, whereas anomalous yet universal super-Poissonian-type scaling $\propto L^{\alpha}$ $(\alpha\sim 2.7$ up to $L=20)$ appears for strong measurement, although the unconditional steady state is always the trivial infinite-temperature state. This crossover results in two distinct regimes in estimating the rate of quantum jumps: for strong (weak) measurement, the uncertainty is enhanced (suppressed) when the system size increases. Strikingly, the crossover is unique to SFQJ and absent for fluctuations in the whole system, where the Poissonian-type scaling always appears. We elucidate that the anomalous scaling stems from the integrated many-body autocorrelation functions for the unconditional dynamics and that the universal exponents are not determined by the Liouvillian gap. Measurement-induced many-body phenomena presented in our work can be accessed in ultracold atom experiments without postselections. See Fig.~\ref{fig_schematic} for the summary of our work.

\textit{Continuous monitoring and SFQJ}--We consider interacting hard-core boson chains of length $L$ ($=2\mathbb N$) with the periodic boundary condition:
\begin{align}
H=\sum_{j=1}^L\frac{J_{xy}}{2}(b_{j+1}^\dag b_j + b_{j+1}b_j^\dag) + \sum_{j=1}^LJ_zn_{j+1}n_j,
\label{eq_HB}
\end{align}
where $b_j$ is the bosonic annihilation operator satisfying the hard-core constraint $b_j^2=0$, and $n_j=b_j^\dag b_j$ is the particle number operator. We call this model the XXZ spin chain as the Hamiltonian \eqref{eq_HB} is exactly mapped to the XXZ model with the anisotropy $J_z/J_{xy}$. Then, we study the dynamics under continuous monitoring of a local particle number by employing the quantum trajectory method \cite{Fuji20, Yamamoto23}. The stochastic Schr\"odinger equation that governs the dynamics reads \cite{Daley14}
\begin{align}
d|\psi(t)\rangle=-iH |\psi(t)\rangle dt +\sum_{j=1}^L \bigg(\frac{n_j|\psi(t)\rangle}{\sqrt{\langle n_j\rangle}}-|\psi(t)\rangle \bigg)dN_j,
\label{eq_stochastic}
\end{align}
where $\langle\cdot\rangle$ denotes a quantum expectation value for the state $|\psi(t)\rangle$. Here, a discrete random variable $dN_j=0,1$ that counts the increment of a jump at site $j$ is chosen according to $dN_jdN_k=\delta_{jk}dN_j$ and $E[dN_j]=\gamma\langle n_j\rangle dt$, where $E[\cdot]$ represents an ensemble average over the stochastic process. In the following, we assume that the initial state is prepared in the N\'eel state $|1010\cdots\rangle$. Importantly, the state ${\rho}(t)$ averaged over the measurement outcomes becomes a maximally mixed state $\rho_\mathrm{ss}= {I}/D_0$ after long times, irrespective of measurement strengths $\gamma\:(>0)$. Here, $I$ is the identity matrix and $D_0$ is the dimension of the Hilbert space in the half-filling sector.

As a primary quantity of interest, we introduce the net number of quantum jumps for the half-chain subsystem [see Fig.~\ref{fig_schematic}(a)], 
\begin{align}
N_\mathrm{jump}^\mathrm{half}(T)=\sum_{j=1}^{L/2} \int_{t_\mathrm{ss}}^{t_\mathrm{ss}+T}d N_j,
\end{align}
where $t_\mathrm{ss}$ is sufficiently large such that $\rho(t_\mathrm{ss})\simeq \rho_\mathrm{ss}$. We especially focus on the variance of $N_\mathrm{jump}^\mathrm{half}(T)$, or SFQJ, which is defined by
\begin{align}
V[N_\mathrm{jump}^\mathrm{half}(T)]=E[N_\mathrm{jump}^\mathrm{half}(T)^2]-E[N_\mathrm{jump}^\mathrm{half}(T)]^2.
\label{eq_SFQJ}
\end{align}
For comparison, we also define the number of quantum jumps for the whole system, $N_\mathrm{jump}^\mathrm{total}(T)=\sum_{j=1}^L\int_{t_\mathrm{ss}}^{t_\mathrm{ss}+T}dN_j$, and consider its variance. Note that the average jump number for $N_\mathrm{jump}^\mathrm{total}(T)$ is given by $E[N_\mathrm{jump}^\mathrm{total}(T)]=\gamma \langle\sum_{j=1}^L n_j\rangle T=\gamma LT/2$, which also leads to $E[N_\mathrm{jump}^\mathrm{half}(T)]=\gamma LT/4$ because of the symmetry. We remark that SFQJ is different from the so-called bipartite fluctuations $\langle n_\mathrm{half}^2\rangle - \langle n_\mathrm{half}\rangle^2$ for $n_\mathrm{half}=\sum_{i=1}^{L/2}n_i$ \cite{Mog23}.

\textit{Measurement-induced crossover}--We first demonstrate the measurement-induced crossover of SFQJ by numerically studying the Heisenberg model ($J_z=1$). As shown in Fig.~\ref{fig_variance}(a), we see that $V[N_\mathrm{jump}^\mathrm{total}(T)]$ exhibits a trivial scaling proportional to $\gamma L$, irrespective of $\gamma$. On the other hand, SFQJ shown in Fig.~\ref{fig_variance}(b) does not behave linearly against $\gamma$ nor $L$, and we find a crossing in $V[N_\mathrm{jump}^\mathrm{half}(T)]/L^2$ around $\gamma_c\sim0.5$ as seen in Fig.~\ref{fig_variance}(c). This indicates a measurement-induced crossover of SFQJ in the system-size scaling, and in fact in Fig.~\ref{fig_variance}(d), we see that the scaling is estimated to show a crossover from $\propto L^{1.13}$ for $\gamma=0.05$ to $\propto L^{2.45}$ for $\gamma=1$. However, we note that universal exponents are different from these values due to the finite-size effect. This is because SFQJ is given as a sum of two terms whose $L$ dependence is different as detailed later, which leads to the deviations of exponents from the universal ones for finite-system sizes. We can conduct the same numerical simulation for XX ($J_z=0$) and XXZ ($J_z=0.5$, $1.5$) models as shown in the Supplemental Material \cite{Suppl}. All these models exhibit the measurement-induced crossover, signifying an anomalous enhancement of SFQJ in many-body quantum systems under continuous monitoring. We also mention that such an anomalous scaling does not exist in the Ising model ($J_{xy}=0$, $J_z=1$) due to particle-number conservation at a single site \cite{Suppl}.

\textit{Relation to unconditional dynamics}--We examine the measurement-induced crossover with the help of the unconditional dynamics of Eq.~\eqref{eq_stochastic} and clarify that SFQJ is composed of two terms: dynamical activity with the volume law that governs SFQJ for sufficiently weak measurement strength, and an integrated autocorrelation function that causes the anomalous enhancement of SFQJ. First of all, we should beware that SFQJ \eqref{eq_SFQJ} is unique to quantum trajectories because measurement outcomes are averaged over all possible sequences in the unconditional dynamics. However, the counting variable is still evaluated by using the Liouvillian, and we can write down SFQJ in terms of the ensemble-averaged dynamics \cite{Landi24}. To start with, we calculate the noise $D_\mathrm{half}(t)=dV[N_\mathrm{jump}^\mathrm{half}(t)]/dt$ for a subsystem in the steady state (see Appendix A in End Matter for the detailed derivation) as
\begin{align}
&D_\mathrm{half}(T)\notag\\
&=K_\mathrm{half}+2\int_0^T d\tau \left\{\mathrm{Tr}[\mathcal J_\mathrm{half} e^{\mathcal L \tau}\mathcal J_\mathrm{half} \rho_\mathrm{ss}]-J_\mathrm{half}^2\right\},
\label{eq_noise_T}
\end{align}
where
$\mathcal L(\rho)\equiv -i[H,\rho]+\gamma \sum_i(-\frac{1}{2}\{n_i,\rho\}+n_i\rho n_i)$ is the Liouvillian superoperator for the ensemble-averaged dynamics of Eq.~\eqref{eq_stochastic}, the subsystem superoperator $\mathcal J_\mathrm{half}$ is defined as $\mathcal J_\mathrm{half}\rho(t)\equiv\gamma\sum_{i=1}^{L/2}n_i\rho(t)n_i$, and $K_\mathrm{half}\equiv\gamma\sum_{i=1}^{L/2}\mathrm{Tr}[n_i\rho_\mathrm{ss}]$ and $J_\mathrm{half}\equiv \mathrm{Tr}[\mathcal J_\mathrm{half}\rho_\mathrm{ss}]$ both reduce to $K_\mathrm{half}=J_\mathrm{half}=\gamma L /4$, which is nothing but (half of) the dynamical activity in the steady state, $E[dN_\mathrm{jump}^\mathrm{total}(t)]/dt= 2E[dN_\mathrm{jump}^\mathrm{half}(t)]/dt=\gamma L/2$. By numerically calculating the long-time dynamics in Eq.~\eqref{eq_noise_T}, we find that the noise becomes constant for large $T$ and does not involve a contribution of $\mathcal O(T)$ \cite{Landi24}. Then, we obtain SFQJ as $V[N_\mathrm{jump}^\mathrm{half}(T)]=D_\mathrm{half}T$, which leads to
\begin{align}
V[N_\mathrm{jump}^\mathrm{half}(T)]&=V_\mathrm{act}(T) + V_\mathrm{anom}(T),\notag\\
&=\frac{\gamma L T}{4}+2\gamma^2 T\int_0^T d\tau \langle n_\mathrm{half}^\prime(\tau)n_\mathrm{half}^\prime\rangle_\infty,
\label{eq_variancehalf_T}
\end{align}
where $n_\mathrm{half}^\prime(\tau)=n_\mathrm{half}(\tau)-\langle n_\mathrm{half}\rangle_\infty$, the Heisenberg picture for the adjoint Liouvillian $\tilde {\mathcal L}(A)\equiv i[H,A]+\gamma \sum_i(-\frac{1}{2}\{n_i,A\}+n_iA n_i)$ is represented as $n_\mathrm{half}(\tau)=e^{\tilde{\mathcal L} \tau}n_\mathrm{half}$, and $\langle \cdot \rangle_\infty$ stands for the expectation value with respect to $\rho_\mathrm{ss}$.
\ccred{This demonstrates that the super-Poissonian-type anomalous enhancement of SFQJ in the measurement-induced crossover originates from the integrated autocorrelation function}
\begin{align}
\mathcal C_\mathrm{auto}^\mathrm{half}\equiv\int_0^T d\tau C_\mathrm{auto}^\mathrm{half}(\tau)\equiv \int_0^T d\tau \langle n_\mathrm{half}^\prime(\tau)n_\mathrm{half}^\prime\rangle_\infty.
\end{align}
Here, the positive (zero) correlation $V[N_\mathrm{jump}^\mathrm{half}(T)]>E[N_\mathrm{jump}^\mathrm{half}(T)]$ ($V[N_\mathrm{jump}^\mathrm{half}(T)]=E[N_\mathrm{jump}^\mathrm{half}(T)]$) is referred to as the super-Poissonian (Poissonian) statistics. We remark that, as the particle number in the whole chain is a conserved quantity, the anomalous term is absent for the whole system, and the variance $V[N_\mathrm{jump}^\mathrm{total}(T)]$ reduces to the dynamical activity, which shows a Poissonian-type volume law irrespective of the measurement strength.

\begin{figure}[t]
\includegraphics[width=8.5cm]{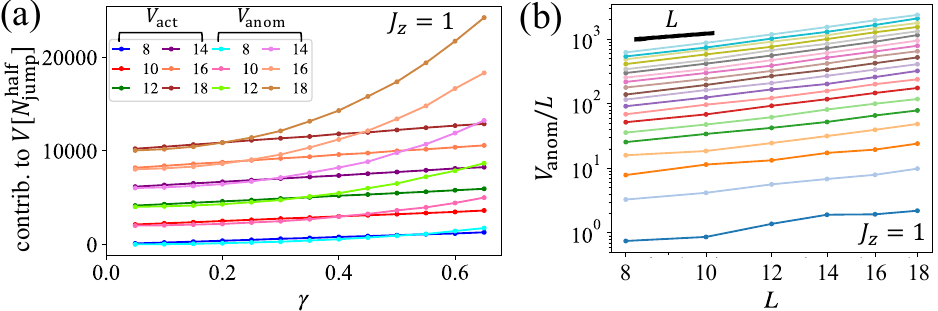}
\caption{Numerical results for $V_\mathrm{anom}$ and $V_\mathrm{act}$ based on the quantum trajectory method for the Heisenberg model. (a) The leading contribution of $V_\mathrm{anom}$ and $V_\mathrm{act}$ changes at a critical measurement strength. (b) System-size scaling of $V_\mathrm{anom}/L$. Data are shifted by $2000$ as $L$ is increased as $L=8, 10, \cdots, 18$ in (a) to improve visualization, and measurement strengths are $\gamma=0.05, 0.1, \cdots, 1$ from bottom to top in (b). The parameters and methods are the same as in Fig.~\ref{fig_variance}.}
\label{fig_variance_anomalous}
\end{figure}

\textit{Origin of the measurement-induced crossover}--Using the quantum trajectory method, we numerically simulate the Heisenberg chain and obtain the contribution of $V_\mathrm{anom}$ and $V_\mathrm{act}$ to SFQJ. \ccred{The two terms $V_\mathrm{anom}$ and $V_\mathrm{act}$ are quantitatively evaluated with the help of the analytical expression \eqref{eq_variancehalf_T} as shown in Fig.~\ref{fig_variance_anomalous}(a), where they cross at the critical measurement strength $\gamma_c$, and the leading contribution to SFQJ changes from $V_\mathrm{act}$ (Poissonian) to $V_\mathrm{anom}$ (super-Poissonian) as $\gamma$ is increased.} \cred{This causes a measurement-induced crossover of SFQJ illustrated in Fig.~\ref{fig_variance}, which offers the novel platform to investigate many-body phenomena under measurement absent in the entire system. We note that the anomalous scaling is characterized by a mechanism distinct from, e.g., measurement-induced entanglement transitions, whose presence depends on interactions in one dimension \cite{Poboiko23}.}

In Fig.~\ref{fig_variance_anomalous}(b) and also in other parameters of $J_z$ \cite{Suppl}, we find by performing the finite-size scaling analysis for $L=8, 10,\cdots, 18$ that 
\begin{align}
V_\mathrm{anom}\propto L^\alpha,
\label{eq_scaling}
\end{align}
where \ccred{$\alpha=2.64\pm0.02, 2.66\pm0.02, 2.67\pm0.02, 2.67\pm0.02$} for $J_z=0, 0.5, 1, 1.5$ with $\gamma=1$, respectively. \ccred{Here, the statistical uncertainty of the scaling exponent arises from averaging over different time intervals.} Remarkably, this indicates that the anomalous scaling of SFQJ is characterized by the universal exponent $\alpha\sim2.7$. Since $V_\mathrm{anom}$ is also proportional to $\gamma^2 c(\gamma)$, where $c(\gamma)$ is an increasing polynomial function of $\gamma$ with $c(0)\ge0$ \cite{Suppl}, SFQJ is dominated by $V_\mathrm{anom}\propto L^\alpha$ for strong $\gamma=\mathcal{O}(L^0)$ instead of $V_\mathrm{act}\propto L$ for weak $\gamma$ [which should be $\mathrm o(L^0)$], the fact of which is the direct origin of the crossover of SFQJ [see Fig.~\ref{fig_schematic}(b)].
\ccred{Notably, the anomalous yet universal super-Poissonian statistics of SFQJ even persists in the thermodynamic limit, further highlighting the significance of the measurement-induced universality (see Appendix B for further evaluation of the rigidity of $\alpha$).}
Physically, such an enhancement of SFQJ is regarded as a kind of Zeno effects \cite{Itano90}, which indicate that, if a quantum jump occurs in a subsystem, the next jump is also likely to occur in the same subsystem.

\ccred{We argue that the global U(1) symmetry without internal particle number conservations is the essential physical factor for the measurement-induced universality. In Appendix B in End Matter, we indeed demonstrate that the anomalous scaling with a quantitatively similar universal exponent at quarter filling and even in nonintegrable models with U(1) symmetry. Note that, if the total particle conservation is broken, the super-Poissonian statistics of the fluctuation would emerge in the whole system as well as the subsystem owing to the Zeno effect.}

\begin{figure}[t]
\includegraphics[width=8.5cm]{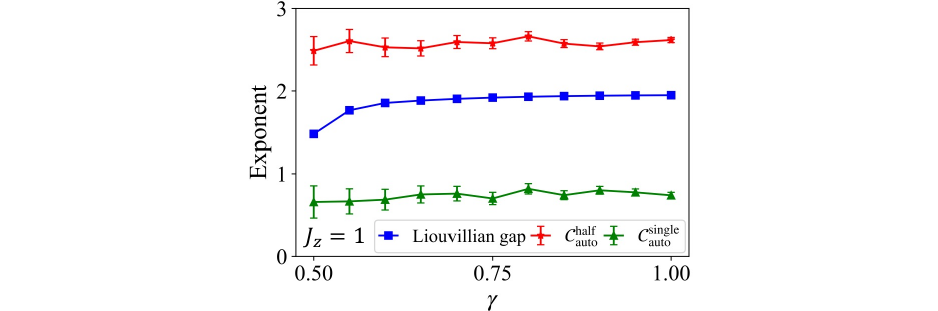}
\caption{Exponents $a$, $b$, and $c$ of the Liouvillian gap $\Delta\propto 1/L^a$ (blue), $\mathcal C_\mathrm{auto}^\mathrm{single}\propto L^b$ (green), and $\mathcal C_\mathrm{auto}^\mathrm{half}\propto L^c$ (red) obtained from the finite-size scaling analysis with $L=8, 10, 12, 14$ for the Heisenberg model. The anomalous scaling of SFQJ cannot be captured by the Liouvillian gap.}
\label{fig_autocorrelation}
\end{figure}

\textit{Impact on the precision}--One interesting consequence of the above anomalous scaling is the drastically enhanced uncertainty in estimating the rate of quantum jumps occurring in the subsystem. Let us estimate the rate of quantum jumps in the subsystem by $\bar{I}_\mathrm{half}(T)=N_\mathrm{jump}^\mathrm{half}(T)/T$ \cite{Landi24}. We then obtain
\begin{gather}
E[\bar{I}_\mathrm{half}(T)]=\frac{E[N_\mathrm{jump}^\mathrm{half}(T)]}{T}=J_\mathrm{half} \propto \gamma L,\\
V[\bar{I}_\mathrm{half}(T)]=\frac{V[N_\mathrm{jump}^\mathrm{half}(T)]}{T^2} = \frac{D_\mathrm{half}}{T} \propto \frac{\gamma L +  \gamma^2c(\gamma) L^{\alpha}}{T}.
\end{gather}
To estimate the true rate $E[\bar{I}_\mathrm{half}(T)]$ accurately, we shall evaluate the estimation uncertainty defined by the (squared) coefficient of variation
\begin{align}
\frac{V[\bar{I}_\mathrm{half}(T)]}{(E[\bar{I}_\mathrm{half}(T)])^2}=\frac{D_\mathrm{half}}{J_\mathrm{half}^2 T} \equiv \frac{T_\mathrm{th}}{T},
\label{eq_ergodicity}
\end{align}
where $T_\mathrm{th}\equiv D_\mathrm{half}/J_\mathrm{half}^2$ is frequently discussed in thermodynamic and kinetic uncertainty relations \cite{Landi24}; here, we rather focus on the $L$-dependence in the large-$T$ dynamics. For sufficiently weak $\gamma$, we find
\begin{align}
T_\mathrm{th}\propto L^{-1}.
\label{eq_poisson}
\end{align}
This means that the error for evaluating $E[\bar{I}_\mathrm{half}(T)]$ diminishes fast enough, and $E[\bar{I}_\mathrm{half}(T)]$ is accurately determined for larger system sizes. On the other hand, for strong measurement satisfying $\gamma\sim\mathcal O(L^0)$, we obtain
\begin{align}
T_\mathrm{th} \propto L^{\alpha-2},
\label{eq_superpoisson}
\end{align}
with $\alpha\sim2.7$ (up to $L=20$), which means that the uncertainty in estimating $E[\bar{I}_\mathrm{half}(T)]$ is enhanced as the system size is increased [see Fig.~\ref{fig_schematic}(b)]. We remark that, as the estimation uncertainty in Eq.~\eqref{eq_superpoisson} is proportional to $L^{\alpha-2}$, particular super-Poissonian-type variance satisfying $\alpha>2$ enhances $T_\mathrm{th}$ with increasing $L$. Moreover, the estimation uncertainty \eqref{eq_ergodicity} is related to the ergodicity of a quantum trajectory for the jump statistics \cite{Cresser01}, i.e., the concept that the long-time average of a single realization yields the ensemble average in almost all trajectories. Equation \eqref{eq_superpoisson} means that the ergodicity of a quantum trajectory tends to be disturbed for larger system sizes in XXZ spin chains under continuous monitoring. We note that the fluctuation for a whole system always follows the trivial Poissonian statistics, whose estimation uncertainty reduces to Eq.~\eqref{eq_poisson} for arbitrary $\gamma$.

\textit{Autocorrelation function and Liouvillian gap}--Finally, it is illustrative to what extent the scaling \eqref{eq_scaling} is affected by the Liouvillian gap $\Delta$, the smallest nonzero value among the real parts of the eigenvalues of $-\mathcal{L}$ (see Appendix C in End Matter). This is because the asymptotic decay of the single-site autocorrelation function is domincated by $\Delta$ as $|C_\mathrm{auto}^\mathrm{single}(\tau)|\equiv|\langle n_1^\prime(\tau)n_1^\prime\rangle_\infty|\sim e^{-\Delta \tau}$ $(\tau\to\infty)$ \cite{Mori23, Shirai24}.
We especially compare the exponent of $\mathcal C_\mathrm{auto}^\mathrm{single}\equiv\int_0^T d\tau C_\mathrm{auto}^\mathrm{single}(\tau)$ with that solely extracted from the Liouvillian gap; $\mathcal C_\mathrm{auto}^\mathrm{single}\stackrel{?}{\sim} \int_0^T e^{-\Delta \tau}\sim 1/\Delta$. In Fig.~\ref{fig_autocorrelation}, we depict the exponent of the inverse Liouvillian gap obtained from the exact diagonalization and that of $\mathcal C_\mathrm{auto}^\mathrm{single}$ calculated from the quantum trajectory method for the Heisenberg model. We find that, the Liouvillian gap estimated from the finite-size scaling analysis reads $\Delta\propto1/L^{1.95}$ for $\gamma=1$, which is close to the analytical result $\Delta\propto1/L^2$ for $\gamma\gg1/L$ \cite{Cai13, Znidaric15}. On the other hand, we obtain \ccred{$\mathcal C_\mathrm{auto}^\mathrm{single}\propto L^{0.74\pm0.04}$} for $\gamma=1$, which indicates that the scaling for the integrated autocorrelation function is largely governed by the transient dynamics \cite{Shirai24} rather than the Liouvillian gap. We emphasize that, as the dynamical activity is proportional to $L^0$ for a single site, the exponent for $\mathcal C_\mathrm{auto}^\mathrm{single}$ is also anomalous.
%
%
As for $\mathcal C_\mathrm{auto}^\mathrm{half}$ shown in Fig.~\ref{fig_autocorrelation}, which reads \ccred{$\mathcal C_\mathrm{auto}^\mathrm{half}\propto L^{2.62\pm0.03}$} for $\gamma=1$ \cite{finitesize}, we find that not only the diagonal elements corresponding to $\mathcal C_\mathrm{auto}^\mathrm{single}$ but also off-diagonal components $\mathcal C_\mathrm{auto}^{ij}\equiv\int_0^T d\tau\langle n_i^\prime(\tau)n_j^\prime\rangle_\infty$ ($i\neq j$) therein contribute to the scaling of $\mathcal C_\mathrm{auto}^\mathrm{half}$ because $\mathcal C_\mathrm{auto}^\mathrm{half}\gg L/2\cdot \mathcal C_\mathrm{auto}^\mathrm{single}$ holds. Thus, we cannot estimate the anomalous scaling exponent for $\mathcal C_\mathrm{auto}^\mathrm{half}$ given in Eq.~\eqref{eq_scaling} by the Liouvillian gap.

\textit{Experimental proposal and conclusions}--Our main discoveries, i.e., the measurement-induced crossover of SFQJ and its anomalous scaling, are accessible in experiments without postselection because the only information we need is times and positions of quantum jumps.
\ccred{For example, in ultracold atoms, hard-core bosons are naturally realized by employing the Tonks-Girardeau gas \cite{Paredes04, Kinoshita04}, and SFQJ is obtained by recording the photon scattering with a probe light \cite{Luschen17, Patil15} and quantum-gas microscopes \cite{Bakr09, Sherson10}. We then realize the measurement-induced crossover by tuning the light intensity, while this may be affected by heating effects as the detection fidelity of scattered photons would be limited.}
We also suggest to change the system size instead by fixing the probe intensity; SFQJ shows the volume law for sufficiently small system sizes, whereas it exhibits anomalous universal scaling $\propto L^\alpha$ ($\alpha\sim2.7$ up to $L=20$) for large system sizes.

We emphasize that the measurement-induced crossover, which is the gradual change of SFQJ at finite system size \cite{Sierant25}, is significant irrespective of its persistence in the thermodynamic limit.
\ccred{This is because crossover phenomenon away from thermodynamic limit (say, order of $10$ or $10^2$), which is relevant for experiments of trapped ions and ultracold atoms in optical lattices, has served as a variety of significant research topics. In open quantum systems, the qualitative difference between weak dissipation of $\gamma=\mathrm o(L^0)$ and strong dissipation of $\gamma=\mathcal O(L^0)$ was studied for, e.g., the instantaneous decay rate of autocorrelation functions \cite{Shirai24} and the computational complexity under noise in random circuit sampling \cite{Morvan24}.}

In this Letter, we have discovered hitherto unnoticed crossover of measurement-induced many-body properties free from postselection, even though the unconditional steady state trivially becomes a maximally mixed state insensitive to measurement strength. To highlight this, we have shown that SFQJ serves as a new indicator for quantum jumps in many-body systems under measurement, in stark contrast to the fluctuation in the whole system. \ccred{Our results gives a new twist for investigating measurement-induced many-body universality in open quantum systems.} The next step would be the evaluation of the universality class and investigation of the impact of dissipative phase transitions \cite{Minganti18} on the anomalous scaling in prototypical many-body models. In addition, it is worth studying the relation  between our results and well-known universal scalings in many-body problems, e.g., long-time tails in hydrodynamics \cite{Bohrdt17}. \cred{Finally, analytically evaluating the universal exponent obtained in our study is an intriguing future problem.}

%


\begin{acknowledgments}
\textit{Acknowledgments}--We are grateful to Igor Lesanovsky, Marko \v{Z}nidari\v{c}, Kazuya Fujimoto, Taiki Ishiyama, and Hironobu Yoshida for fruitful discussions.
This work was supported by JSPS Program for Forming Japan's Peak Research Universities (J-PEAKS) Grant No.~JPJS00420230008, JST ERATO Grant No.\ JPMJER2302, and KAKENHI Grants No.\ JP24K16982, and No.\ JP25K17327.
This work was also partly supported by Murata Science and Education Foundation, Hirose Foundation, Precise Measurement Technology Promotion Foundation, Fujikura Foundation, Toyota Riken Scholar Program, and Support Center for Advanced Telecommunications Technology Research. The numerical calculations were partly carried out with the help of QuSpin \cite{Weinberg17}. The simulations were performed in part with the help of the supercomputing system in ISSP, the University of Tokyo. K.Y. thanks Juntaro Fujii and Soma Takemori for their helpful supports concerning the usage of the supercomputing system.
\end{acknowledgments}

\textit{Data availability}--The data that support the findings of this Letter are openly available \cite{Zenodo}.

\nocite{apsrev42Control}
\bibliographystyle{apsrev4-2}
\bibliography{MIPTJumpStatistics.bib}

\begin{thebibliography}{110}
\bibitem [{ste()}]{steadystate}%
  \BibitemOpen
  \href@noop {} {}\bibinfo {note} {\cred{In our calculation, we take
  $t_{\mathrm{ss}}=200$ as one prototypical timescale that makes
  $\rho_{\mathrm{ss}}$ reach a maximally mixed state. Practically, this value
  of $t_{\mathrm{ss}}$ has been chosen such that the local particle number
  relaxes to the infinite-temperature values and the trajectory average of the
  pure-state entanglement entropy saturates to the steady-state
  values.}}\BibitemShut {Stop}%
\bibitem [{fin()}]{finitesize}%
  \BibitemOpen
  \href@noop {} {}\bibinfo {note} {The exponent \ccred{$2.62\pm0.03$} deviates
  a little from the exponent \ccred{$\alpha=2.67\pm0.02$} in
  Eq.~\eqref{eq_scaling} for the Heisenberg model. This stems from the
  system-size difference used in the calculation, where the latter exponent is
  obtained from the finite-size scaling analysis for $L=8, \cdots, 18$ in
  Fig.~\ref{fig_variance_anomalous}(b), while the former one for $L=8, \cdots,
  14$ in Fig.~\ref{fig_autocorrelation}. In the former calculation, we need to
  compare it with the exponent of the Liouvillian gap, where the exact
  diagonalization of the Liouvillian numerically limits the system
  size.}\BibitemShut {Stop}%
\end{thebibliography}


\renewcommand{\theequation}{A\arabic{equation}}
\setcounter{equation}{0}

\onecolumngrid
\appendix

\begin{center}
   \textbf{End Matter}
 \end{center}
 
\twocolumngrid
\cred{\textit{Appendix A: Derivation of Eq.~(\ref{eq_noise_T})}--The noise given in Eq.~\eqref{eq_noise_T} is calculated with the help of the dynamics described by Eq.~\eqref{eq_stochastic}, which is the stochastic Schr\"odinger equation that is mainly used for numerical simulations \cite{Landi24}. To prove it, we first calculate the two-point correlation function
\begin{align}
F(t, t^\prime)=E[I_\mathrm{half}(t)I_\mathrm{half}(t^\prime)]-E[I_\mathrm{half}(t)]E[I_\mathrm{half}(t^\prime)],
\label{eq_funvar}
\end{align}
where $I_\mathrm{half}(t)=dN_\mathrm{jump}^\mathrm{half}(t)/dt$ and $F(t, t+\tau)=F(t+\tau, t)$ is satisfied for $\tau>0$. To evaluate the first term in $F(t, t+\tau)$, we need to calculate the joint probability $P[dN_i(t)=1, dN_j(t+\tau)=1]$, which represents that we have quantum jumps at times $t$ and $t+\tau$ without any other restrictions in between. This is because the following is satisfied:
\begin{align}
E&[I_\mathrm{half}(t)I_\mathrm{half}(t+\tau)]\notag\\
&=\frac{1}{dt^2}\sum_{i,j}^{L/2} P[dN_i(t)=1, dN_j(t+\tau)=1]\notag\\
&=\frac{1}{dt^2}\sum_{i,j}^{L/2} E[P[dN_j(t+\tau)=1|dN_i(t)=1,\rho_c(t)]p_i^c(t)],
\label{eq_E_II}
\end{align}
where we have introduced the conditional probability $P[\:\cdot\:|\:\cdot\:]$ by using $p_i^c(t)\equiv \mathrm{Tr}[\mathcal L_i \rho_c(t)]dt$ with the conditional density matrix $\rho_c(t)$ on a particular sequence of quantum jumps that satisfies $E[\rho_c(t)]=\rho(t)$. Here, $\mathcal L_i$ is the superoperator defined as $\mathcal L_i (\rho)\equiv \gamma n_i\rho n_i$. Note that $p_i^c(t)$ stands for the probability that a quantum jump occurs at time $t$ for $\rho_c(t)$. Equation \eqref{eq_stochastic} indicates that, if a quantum jump occurs at site $i$, the conditional density matrix of the system $\rho_c(t)$ is updated as
\begin{align}
\rho^\prime_c(t)\equiv\frac{\mathcal L_i\rho_c(t)}{\mathrm{Tr}[\mathcal L_i\rho_c(t)]}=\frac{\mathcal L_i\rho_c(t) dt}{p_i^c(t)}.
\label{eq_update_i}
\end{align}
By using the ensemble-averaged dynamics of Eq.~\eqref{eq_stochastic},
\begin{figure}[b]
\includegraphics[width=8.5cm]{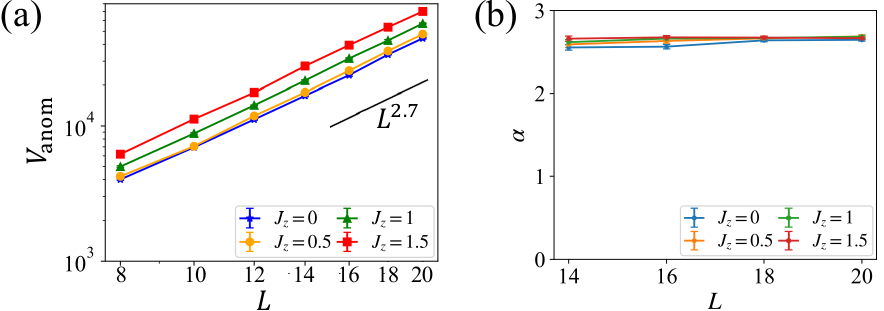}
\caption{\ccred{(a) System-size scaling of $V_\mathrm{anom}$ and (b) the exponent $\alpha$, up to $L=20$ for $\gamma=1$ at half filling. We find the universal scaling $\propto L^\alpha$ ($\alpha\sim2.7$) irrespective of the values of $J_z$.}}
\label{fig_variance_anomalous_L20}
\end{figure}
we obtain the probability that a quantum jump occurs at time $t+\tau$ with respect to the time-evolved state $e^{\mathcal L \tau}\rho^\prime_c(t)$ as
\begin{align}
P&[dN_j(t+\tau)=1|dN_i(t)=1,\rho_c(t)]\notag\\
&=\mathrm{Tr}[\mathcal L_j e^{\mathcal L \tau}\rho^\prime_c(t)]dt=\frac{(dt)^2}{p_i^c(t)}\mathrm{Tr}[\mathcal L_j e^{\mathcal L \tau}\mathcal L_i\rho_c(t)].
\label{eq_E_II2}
\end{align}
Then, Eq.~\eqref{eq_E_II} is calculated with the help of Eq.~\eqref{eq_E_II2} as
\begin{align}
E[I_\mathrm{half}(t)I_\mathrm{half}(t+\tau)]=\mathrm{Tr}[\mathcal J_\mathrm{half} e^{\mathcal L \tau}\mathcal J_\mathrm{half} \rho(t)].
\end{align}
The case $\tau=0$ is treated separately, leading to
\begin{align}
E[I_\mathrm{half}(t)^2]=\frac{1}{(dt)^2}\sum_{i=1}^{L/2}E[dN_i(t)]=\frac{1}{dt}K_{\mathrm{half}}(t),
\end{align}
which results in
\begin{align}
F(t, t+\tau)=&\delta(\tau)K_\mathrm{half}(t)+\mathrm{Tr}[\mathcal J_\mathrm{half} e^{\mathcal L \tau}\mathcal J_\mathrm{half}\rho(t)]\notag\\
&-J_\mathrm{half}(t)J_\mathrm{half}(t+\tau).
\label{eq_fluctuation}
\end{align}
Here, we have introduced $K_\mathrm{half}\equiv\gamma\sum_{i=1}^{L/2}\mathrm{Tr}[n_i\rho(t)]$.
As SFQJ \eqref{eq_SFQJ} is rewritten by using Eq.~\eqref{eq_funvar} as
\begin{align}
V[N_\mathrm{jump}^\mathrm{half}(T)]=\int_{t_\mathrm{ss}}^{T+t_\mathrm{ss}}dt \int_{t_\mathrm{ss}}^{T+t_\mathrm{ss}}dt^\prime F(t,t^\prime),
\end{align}
we find that the noise $D_\mathrm{half}(t)=dV[N_\mathrm{jump}^\mathrm{half}(t)]/dt$ is given by
\begin{align}
D_\mathrm{half}(T)&=2\int_{t_\mathrm{ss}}^{T+t_\mathrm{ss}}dt^\prime F(T+t_\mathrm{ss},t^\prime),\notag\\
&=2\int_0^Td\tau F(T+t_\mathrm{ss}-\tau, T+t_\mathrm{ss}).
\label{eq_noise0T_app}
\end{align}
Finally, by substituting Eq.~\eqref{eq_fluctuation} into \eqref{eq_noise0T_app}, we obtain
\begin{align}
D_\mathrm{half}&(T)=2\int_0^T d\tau \delta(\tau)K_\mathrm{half}(T+t_\mathrm{ss}-\tau)\notag\\
+&2\int_0^T d\tau\Big\{\mathrm{Tr}[\mathcal J_\mathrm{half} e^{\mathcal L \tau}\mathcal J_\mathrm{half}\rho(T+t_\mathrm{ss}-\tau)]\notag\\
-&J_\mathrm{half}(T+t_\mathrm{ss}-\tau)J_\mathrm{half}(T+t_\mathrm{ss})\Big\},
\label{eq_noise_T_app}
\end{align}
which reduces to Eq.~\eqref{eq_noise_T} by replacing $K_{\mathrm{half}}(T+t_{\mathrm{ss}}-\tau)$, $J_{\mathrm{half}}(T+t_{\mathrm{ss}}-\tau)$, $J_{\mathrm{half}}(T+t_{\mathrm{ss}})$, and $\rho(T+t_{\mathrm{ss}}-\tau)$ with their steady-state values.}

\begin{figure}[t]
\includegraphics[width=8.5cm]{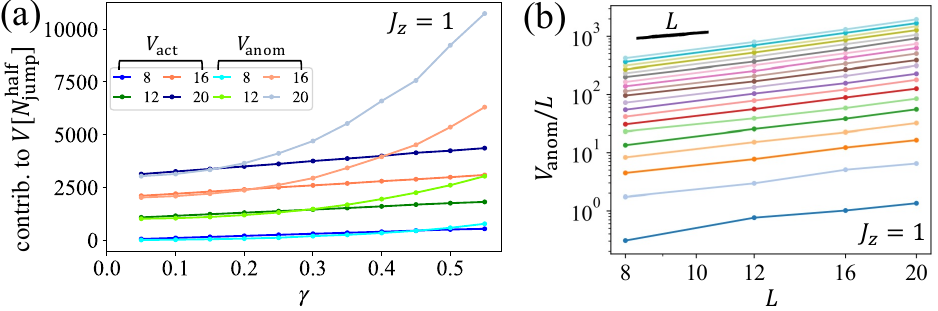}
\caption{\cred{Numerical results for $V_\mathrm{anom}$ and $V_\mathrm{act}$ based on the quantum trajectory method for the Heisenberg model at quarter filling. (a) The leading contribution of $V_\mathrm{anom}$ and $V_\mathrm{act}$ changes at a critical measurement strength similar to the half-filling case. (b) System-size scaling of $V_\mathrm{anom}/L$. Data are shifted by $1000$ as $L$ is increased as $L=8, 12, 16, 20$ in (a) to improve visualization, and measurement strengths are $\gamma=0.05, 0.1, \cdots, 1$ from bottom to top in (b). The other parameters and methods are the same as in Fig.~\ref{fig_variance}.}}
\label{fig_variance_anomalous_quarter}
\end{figure}

\textit{Appendix B: Numerical results for large system sizes and different initial states}--To further confirm the universality of the super-Poissonian statistics of SFQJ, we perform numerical simulations under continuous monitoring for large system sizes and also change the initial condition to be at quarter filling $\psi(0)=|100010\cdots\rangle$.  First, we calculate SFQJ for $L=20$ at half filling as shown in Figs.~\ref{fig_variance_anomalous_L20}(a),\;(b). By performing the finite-size scaling analysis for $L=8, 10, \cdots,20$, we obtain the scaling exponents for  
$V_\mathrm{anom}\propto L^\alpha$ as \ccred{$\alpha=2.65\pm0.02,2.69\pm0.02,2.68\pm0.02,2.66\pm0.02$} for $J_z=0, 0.5, 1, 1.5$ with $\gamma=1$, respectively. These values are almost the same as the results calculated from the finite-size scaling up to $L=18$ obtained in the main text and support the universality of the exponent.

\cred{Then, we perform the numerical simulation for the quarter-filling case. We remark that, if the system is at quarter filling, we have to change the rate of quantum jumps properly during numerical calculations. Accordingly, $\rho_{\mathrm{ss}}$ should be described in the quarter-filling sector with the use of its Hilbert space dimension $D_0^\prime$, and the first term in Eqs.~\eqref{eq_noise_T} and \eqref{eq_variancehalf_T} is modified since quarter-filling condition gives $E[N_\mathrm{jump}^\mathrm{half}(T)]=\gamma LT/8$ instead of $\gamma LT/4$. As shown in Fig.~\ref{fig_variance_anomalous_quarter}(a), we find the measurement-induced crossover characterized by the crossing of $V_\mathrm{anom}$ and $V_\mathrm{act}$, where the super-Poissonian statistics of SFQJ emerges for strong measurement.}

Moreover, by performing the finite-size scaling analysis for $L=8, 12, 16, 20$ as illustrated in Fig.~\ref{fig_variance_anomalous_quarter}(b), we obtain the anomalous scaling of SFQJ as $V_\mathrm{anom}\propto L^\alpha$, where \ccred{$\alpha=2.69\pm0.02, 2.70\pm0.02, 2.69\pm0.02, 2.72\pm0.02$} for $J_z=0, 0.5, 1, 1.5$ with $\gamma=1$, respectively. Remarkably, these values are quantitatively similar to the exponents obtained for the half-filling condition and suggest that the anomalous scaling of SFQJ exhibits the universal behavior that does not depend on initial conditions.

\begin{figure}[t]
\includegraphics[width=8.5cm]{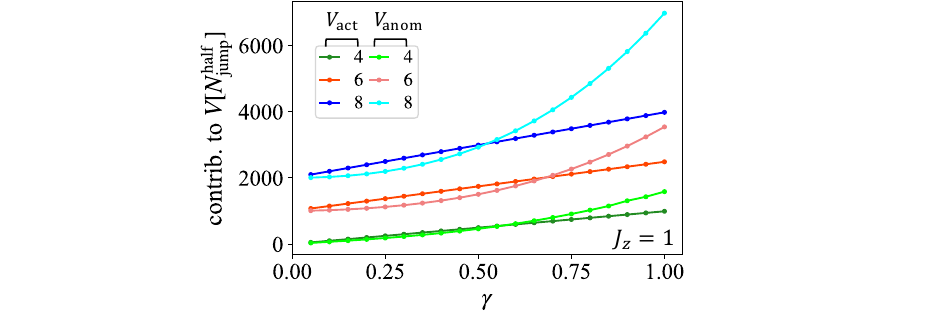}
\caption{\cred{$V_\mathrm{anom}$ and $V_\mathrm{act}$ for the Heisenberg model for system sizes $L=4, 6,8$ obtained with the use of Eq.~\eqref{eq_variance_spc}. The leading contribution of $V_\mathrm{anom}$ and $V_\mathrm{act}$ changes at a critical measurement strength consistent with Fig.~\ref{fig_variance_anomalous}(a). We set $T = 990$ and have numerically checked that $C_\mathrm{auto}^\mathrm{half}$ becomes almost constant for large $T$. Data are shifted by $1000$ as $L$ is increased.}}
\label{fig_autocorrelation_end}
\end{figure}

\ccred{Furthermore, we have tested the robustness of the universality under a global U(1) symmetry for nonintegrable models, by incorporating next nearest neighbor interactions $H_{\mathrm{NN}}=\sum_{j=1}^L J_z^\prime n_{j+2}n_j$ to the Hamiltonian \eqref{eq_HB}. Notably, we obtain $\alpha=2.70\pm0.02,2.70\pm0.02,2.71\pm0.02,2.66\pm0.02$ ($J_z^\prime=0.5$) and $\alpha=2.69\pm0.01,2.72\pm0.02,2.71\pm0.02,2.68\pm0.02$ ($J_z^\prime=1$), up to $L=20$ with $\gamma=1$ for $J_z=0, 0.5, 1, 1.5$, respectively. This result demonstrates that the universality is robust even for nonintegrable Hamiltonians under measurement, going beyond the integrable XX and XXZ Hamiltonians.}

\ccred{Finally, we have tried to extrapolate the exponent to the thermodynamic limit, where we have obtained $\alpha=2.90\pm0.09,2.91\pm0.10,2.82\pm0.08,2.66\pm0.08$ ($J_z^\prime=0$), $\alpha=2.85\pm0.10,2.94\pm0.10,2.83\pm0.09,2.79\pm0.09$ ($J_z^\prime=0.5$), and $\alpha=2.81\pm0.08,2.86\pm0.09,3.02\pm0.09,2.81\pm0.10$ ($J_z^\prime=1$), for $J_z=0,0.5,1,1.5$, respectively. Then, we see that almost all parameter sets (10 out of 12) take the universal value $\alpha=2.85$ within the error bar in the thermodynamic limit. We note that the statistical error is rather large due to the limitation of samples, and this causes a slight deviation for two sets of parameters from $\alpha=2.85$. Thus, though the exact value may slightly deviate to a larger value, this fact still supports the universality of the exponent.}


\renewcommand{\theequation}{C\arabic{equation}}
\setcounter{equation}{0}

\cred{\textit{Appendix C: Spectral decomposition of the Liouvillian}--We explain the spectral decomposition of the Liouvillian for the unconditional dynamics of Eq.~\eqref{eq_stochastic} for calculating the Liouvillian gap as well as $V_\mathrm{anom}$ and $V_\mathrm{act}$, assuming the diagonalizability of $\mathcal{L}$. By employing the basis transformation to the doubled Hilbert space for the density matrix $\rho=\sum_{i,j} \rho_{ij}|i\rangle \langle j |$ as $|i\rangle \langle j |\mapsto |ij)\coloneqq|i\rangle\otimes|j\rangle \in \mathcal H \otimes \mathcal H$ \cite{Yoshioka19, Shibata19, Yamamoto21, Landi24}, Eq.~\eqref{eq_variancehalf_T} is rewritten as
\begin{align}
&V[N_\mathrm{jump}^\mathrm{half}(T)]\notag\\
&=\frac{\gamma LT}{4}+\frac{2\gamma^2 T}{D_0} \sum_{j\neq0}\frac{e^{\lambda_j T}-1}{\lambda_j}(n_\mathrm{half}^\prime|\rho_j^R)(\rho_j^L|n_\mathrm{half}^\prime),
\label{eq_variance_spc}
\end{align}
where the spectral decomposition of the Liouvillian is given by $\mathcal L =\sum_{j\neq0}\lambda_j|\rho_j^R)(\rho_j^L|$, the inner product is defined as $(A|B)=\mathrm{Tr}[A^\dag B]$, and we have restricted both the bra and the ket space to a half-filling sector. By sorting the eigenvalues of $\mathcal L$ to satisfy $0=\lambda_0>\mathrm{Re}[\lambda_1]\ge\mathrm{Re}[\lambda_2]\ge\cdots\ge\mathrm{Re}[\lambda_{D_0^2-1}]$, where we confirm that the unconditional steady state is unique, the Liouvillian gap $\Delta$ that dominates the asymptotic decay rate \cite{Mori20, Haga21} is extracted from the largest real part of the eigenvalue except for $\lambda_0$ as $\Delta = -\mathrm{Re}\lambda_1$. As the Lindblad equation is invariant under the transformation by a real constant in the particle-number jump operator \cite{Breur02}, we refer to the Liouvillian gap in the Heisenberg model with dephasing \cite{Cai13, Znidaric15}. In addition, as shown in Fig.~\ref{fig_autocorrelation_end}, we obtain $V_\mathrm{anom}$ and $V_\mathrm{act}$ with the use of Eq.~\eqref{eq_variance_spc}. Clearly, we find a crossing of $V_\mathrm{anom}$ and $V_\mathrm{act}$ and the critical $\gamma_c$ quantitatively agrees well with that in Fig.~\ref{fig_variance_anomalous}(a) for $L=8$.}


\clearpage

\renewcommand{\thesection}{S\arabic{section}}
\renewcommand{\theequation}{S\arabic{equation}}
\setcounter{equation}{0}
\renewcommand{\thefigure}{S\arabic{figure}}
\setcounter{figure}{0}

\onecolumngrid
\appendix
\begin{center}
\large{Supplemental Material for}\\
\textbf{``Measurement-Induced Crossover of Quantum Jump Statistics in Postselection-Free Many-Body Dynamics"}
\end{center}

\author{Kazuki Yamamoto}
\email{yamamoto@phys.sci.isct.ac.jp}
\affiliation{Department of Physics, Institute of Science Tokyo, Meguro, Tokyo 152-8551, Japan}
\altaffiliation{Former Tokyo Institute of Technology. From October 2024, Tokyo Institute of Technology and Tokyo Medical and Dental University have merged and become Institute of Science Tokyo.}
\author{Ryusuke Hamazaki}
\affiliation{Nonequilibrium Quantum Statistical Mechanics RIKEN Hakubi Research Team, RIKEN Cluster for Pioneering Research (CPR), RIKEN iTHEMS, Wako, Saitama 351-0198, Japan}

\section{Numerical results for subsystem fluctuations of quantum jumps in other many-body models}
\begin{figure*}[h]
\includegraphics[width=18cm]{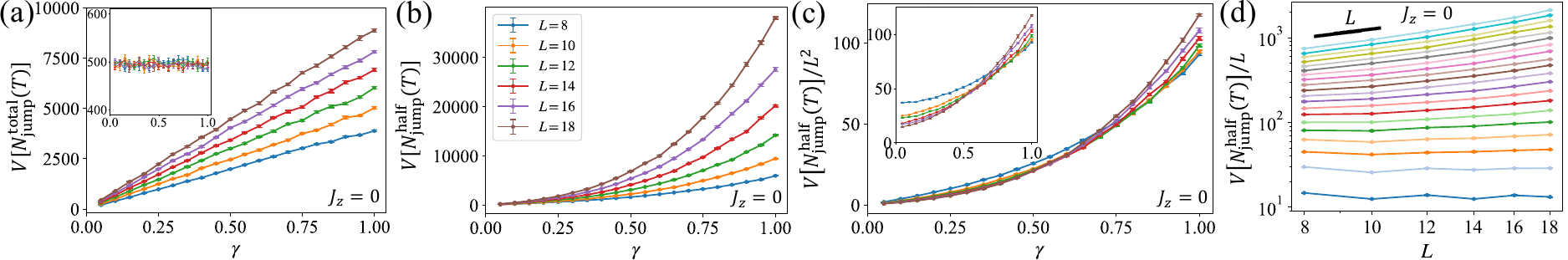}
\caption{Numerical results for the variance of quantum jumps to demonstrate the measurement-induced crossover of SFQJ in the XX model for $832$ trajectories and $T=990$. Data are plotted against $\gamma$ for system sizes $L=8, 10, \cdots, 18$ in (a)-(c) and against $L$ for measurement strengths $\gamma=0.05, 0.1, \cdots, 0.95, 1$ from bottom to top in (d). (a) Variance of net quantum jumps in the whole system [Inset: $V[N_\mathrm{jump}^\mathrm{total}(T)]/(\gamma L)$], (b) SFQJ, (c) $V[N_\mathrm{jump}^\mathrm{half}(T)]/L^2$ [Inset: $V[N_\mathrm{jump}^\mathrm{half}(T)]/(\gamma L^2)$], and (d) $V[N_\mathrm{jump}^\mathrm{half}(T)]/L$. We take the average over 20 time intervals for $t\in [200, 1190], [1190, 2180], \cdots, [19010, 2\times10^4]$.}
\label{fig_varianceXX}
\end{figure*}
We give additional numerical results for the variance of quantum jumps in several many-body models including XX ($J_{xy}=1$, $J_z=0$), XXZ ($J_{xy}=1$, $J_z=0.5, 1.5$), and Ising ($J_{xy}=0$, $J_z=1$) models. We note that, as the dissipator in the Liouvillian cannot be expressed in terms of quadratic operators, we shall include the XX model with dephasing as part of many-body models \cite{Yamamoto20, Haga23, Ishiyama25}. In Fig.~\ref{fig_varianceXX}, we show the variance for the XX chain obtained from the quantum trajectory method. In Fig.~\ref{fig_varianceXX}(a), as expected, the variance for the whole chain does not exhibit the nontrivial behavior and follows the scaling given by $V[N_\mathrm{jump}^\mathrm{total}(T)]=\gamma L T/2$. On the other hand, we find that the subsystem fluctuation of quantum jumps (SFQJ) indicates an anomalous scaling different from the volume law as shown in Fig.~\ref{fig_varianceXX}(b). Then, we see in Fig.~\ref{fig_varianceXX}(c) that $V[N_\mathrm{jump}^\mathrm{half}(T)]/L^2$ shows a crossing around $\gamma_c\sim0.7$ as the measurement rate $\gamma$ is increased. However, the crossing point is rather disturbed compared to the one in Fig.~\ref{fig_variance}(c) in the main text. This comes from the fact that SFQJ in the XX model for weak measurement strength exhibits the stepwise behavior [see the inset in Fig.~\ref{fig_varianceXX}(c)], which may come from the specific integrability of the noninteracting XX spin chain in the unitary limit. Accordingly, we see an oscillation in $V[N_\mathrm{jump}^\mathrm{half}(T)]/L$ for weak measurement strength as shown in Fig.~\ref{fig_varianceXX}(d). It is also worth noting that this oscillation comes from $V_\mathrm{anom}$ that reflects nontrivial subsystem fluctuations. Moreover, we find that measurement-induced crossover of SFQJ emerges in Fig.~\ref{fig_varianceXX}(d), where the scaling is estimated as $V[N_\mathrm{jump}^\mathrm{half}(T)]\propto L^{2.40\:(0.96)}$ for $\gamma=1\:(0.05)$. We note that, for large $L$, these exponents reduce to the pure ones for $V_\mathrm{anom}$ coming from integrated autocorrelation functions. However, for the finite size, they include the correction coming from $V_\mathrm{act}$.

\begin{figure*}[h]
\includegraphics[width=18cm]{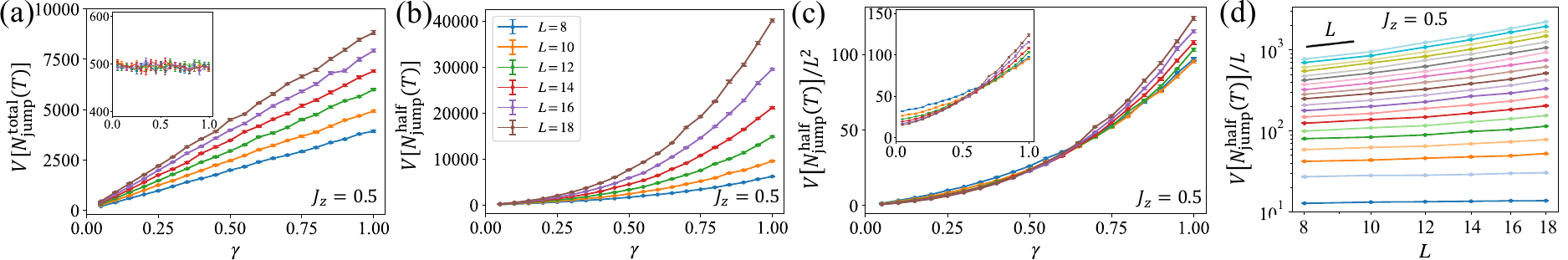}
\caption{Numerical results for the variance of quantum jumps to demonstrate the measurement-induced crossover of SFQJ in the XXZ ($J_z=0.5$) model for $832$ trajectories and $T=990$. Data are plotted against $\gamma$ for system sizes $L=8, 10, \cdots, 18$ in (a)-(c) and against $L$ for measurement strengths $\gamma=0.05, 0.1, \cdots, 0.95, 1$ from bottom to top in (d). (a) Variance of net quantum jumps in the whole system [Inset: $V[N_\mathrm{jump}^\mathrm{total}(T)]/(\gamma L)$], (b) SFQJ, (c) $V[N_\mathrm{jump}^\mathrm{half}(T)]/L^2$ [Inset: $V[N_\mathrm{jump}^\mathrm{half}(T)]/(\gamma L^2)$], and (d) $V[N_\mathrm{jump}^\mathrm{half}(T)]/L$. We take the average over 20 time intervals for $t\in [200, 1190], [1190, 2180], \cdots, [19010, 2\times10^4]$.}
\label{fig_varianceXXZ0p5}
\end{figure*}
\begin{figure*}[h]
\includegraphics[width=18cm]{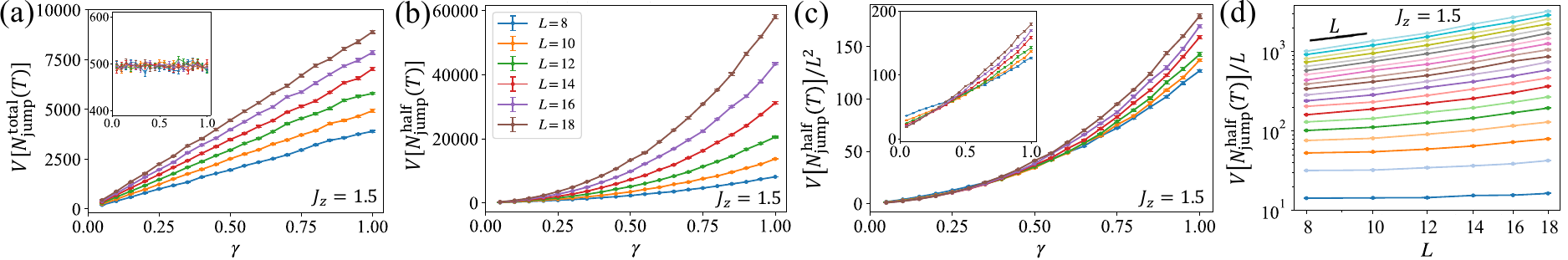}
\caption{Numerical results for the variance of quantum jumps to demonstrate the measurement-induced crossover of SFQJ in the XXZ ($J_z=1.5$) model for $832$ trajectories and $T=990$. Data are plotted against $\gamma$ for system sizes $L=8, 10, \cdots, 18$ in (a)-(c) and against $L$ for measurement strengths $\gamma=0.05, 0.1, \cdots, 0.95, 1$ from bottom to top in (d). (a) Variance of net quantum jumps in the whole system [Inset: $V[N_\mathrm{jump}^\mathrm{total}(T)]/(\gamma L)$], (b) SFQJ, (c) $V[N_\mathrm{jump}^\mathrm{half}(T)]/L^2$ [Inset: $V[N_\mathrm{jump}^\mathrm{half}(T)]/(\gamma L^2)$], and (d) $V[N_\mathrm{jump}^\mathrm{half}(T)]/L$. We take the average over 20 time intervals for $t\in [200, 1190], [1190, 2180], \cdots, [19010, 2\times10^4]$.}
\label{fig_varianceXXZ1p5}
\end{figure*}
In Figs.~\ref{fig_varianceXXZ0p5} and \ref{fig_varianceXXZ1p5}, we depict for the XXZ model ($J_z=0.5,1.5$) the same quantities as those in Fig.~\ref{fig_variance} in the main text and in Fig.~\ref{fig_varianceXX}. We find that the behavior is similar to that of the Heisenberg model shown in Fig.~\ref{fig_variance} in the main text, and the measurement-induced crossover of SFQJ is detected. We find a crossing of $V[N_\mathrm{jump}^\mathrm{half}(T)]/L^2$ around $\gamma_c\sim0.6$ for $J_z=0.5$ and $\gamma_c\sim0.35$ for $J_z=1.5$ in Figs.~\ref{fig_varianceXXZ0p5} and \ref{fig_varianceXXZ1p5}, respectively. Also, we estimate the scaling of SFQJ as $V[N_\mathrm{jump}^\mathrm{half}(T)]\propto L^{2.43\:(1.08)}$ for $\gamma=1\:(0.05)$ for $J_z=0.5$ and $V[N_\mathrm{jump}^\mathrm{half}(T)]\propto L^{2.49\:(1.20)}$ for $\gamma=1\:(0.05)$ for $J_z=1.5$, but the exponents include the finite-size contribution from $V_\mathrm{act}$ as in the case of the XX model.

\begin{figure*}[h]
\includegraphics[width=18cm]{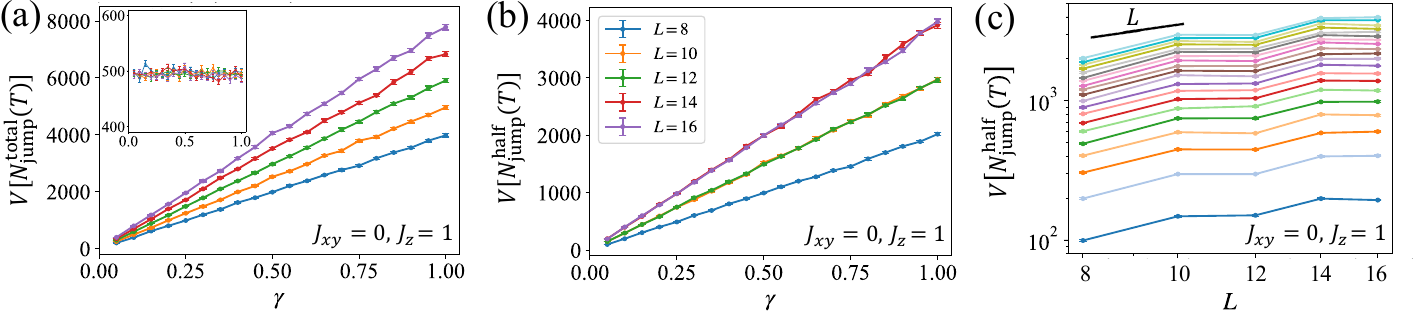}
\caption{Numerical results for the variance of quantum jumps in the Ising model for $832$ trajectories and $T=990$, where no measurement-induced crossover is found even for SFQJ. Data are plotted against $\gamma$ for system sizes $L=8, 10, \cdots, 16$ in (a) and (b) and against $L$ for measurement strengths $\gamma=0.05, 0.1, \cdots, 0.95, 1$ from bottom to top in (c). (a) Variance of net quantum jumps in the whole system [Inset: $V[N_\mathrm{jump}^\mathrm{total}(T)]/(\gamma L)$] and (b), (c) SFQJ. We take the average over 20 time intervals for $t\in [200, 1190], [1190, 2180], \cdots, [19010, 2\times10^4]$.}
\label{fig_varianceZZ}
\end{figure*}
We also give numerical results for the Ising model, which is a trivial yet definite example that does not exhibit the measurement-induced crossover of SFQJ. As shown in Figs.~\ref{fig_varianceZZ}(a) and (b), we find nontrivial behavior in the variance neither for the total system nor the subsystem. Though the scaling of SFQJ shown in Fig.~\ref{fig_varianceZZ}(b) seems to be different from the one of $V[N_\mathrm{jump}^\mathrm{total}(T)]/2=\gamma L T/4$, they agree with each other in the thermodynamic limit. In fact, the stepwise behavior of SFQJ in Fig.~\ref{fig_varianceZZ}(c) reflects the particle-number conservation at a single site in the Ising chain. As the particle number for the half chain is given by $L/4$ $(L/2:\mathrm{even})$ and $L/4+1/2$ $(L/2:\mathrm{odd})$ reflecting the initial N\'eel state, they agree in the thermodynamic limit, and we cannot distinguish the scaling of SFQJ from the one of the total system. We remark that $V_\mathrm{anom}$ becomes zero because the particle number conservation at a single site immediately gives zero autocorrelation in Eq.~\eqref{eq_variancehalf_T} in the main text.

\section{Numerical results for $V_\mathrm{anom}$ and $V_\mathrm{act}$ in other many-body models}
\begin{figure}[h]
\includegraphics[width=18cm]{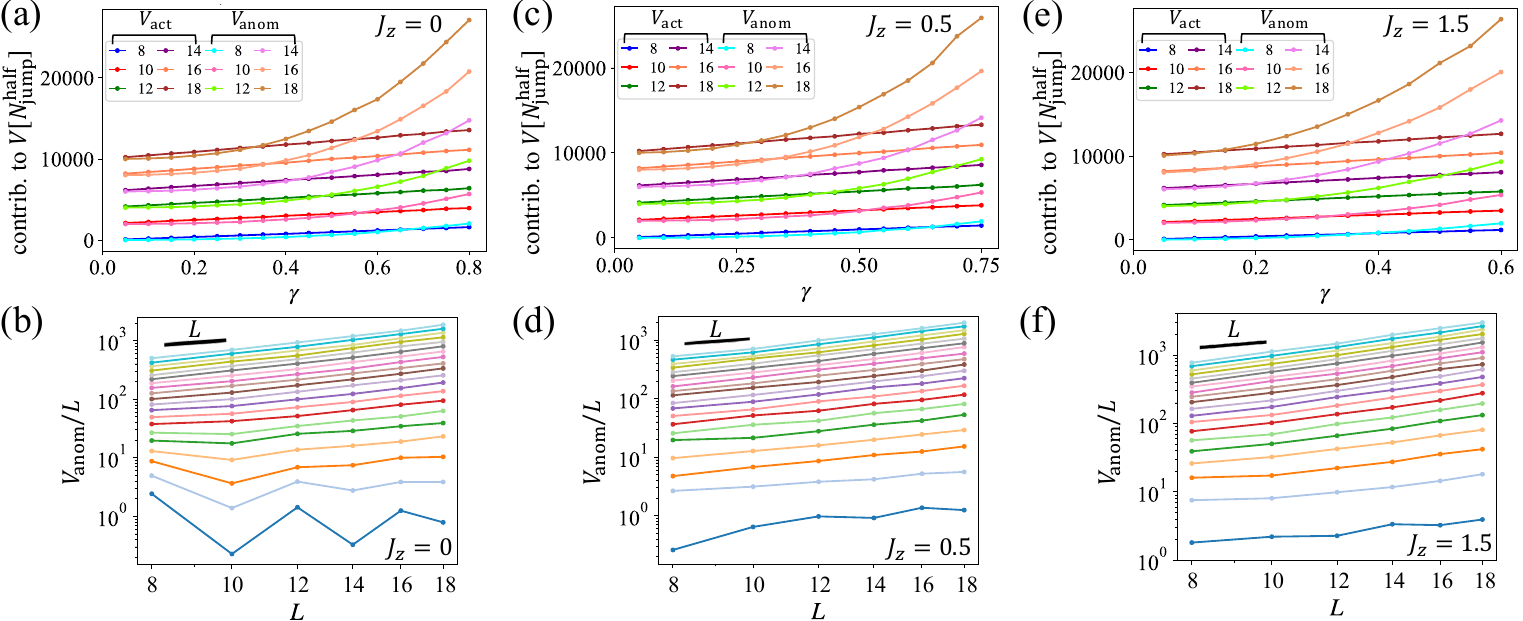}
\caption{Numerical results for $V_\mathrm{anom}$ and $V_\mathrm{act}$ based on the quantum trajectory method for (a), (b) XX, (c), (d) XXZ ($J_z=0.5$), and (e), (f) XXZ ($J_z=1.5$) models. Data are shifted by $2000$ as $L$ is increased as $L=8, 10, \cdots, 18$ to improve visualization in (a),(c), and (e), and measurement strengths are $\gamma=0.05, 0.1, \cdots, 0.95, 1$ from bottom to top in (b), (d), and (f). (a), (c), (e) The leading contribution of $V_\mathrm{anom}$ and $V_\mathrm{act}$ changes at a critical measurement strength. (b), (d), (f) System-size scaling of $V_\mathrm{anom}/L$. The parameters and methods are the same as in Figs.~\ref{fig_varianceXX}-\ref{fig_varianceXXZ1p5}.}
\label{fig_variance_anomalousXXZ}
\end{figure}
In Figs.~\ref{fig_variance_anomalousXXZ}(a), (c), and (e), we show $V_\mathrm{anom}$ and $V_\mathrm{act}$ for XX ($J_z=0$) and XXZ ($J_z=0.5, 1.5$) models obtained from the quantum trajectory method. We find that the dominant contribution of SFQJ changes from $V_\mathrm{act}\propto L$ (Poissonian statistics) to $V_\mathrm{anom}\propto L^\alpha$ (super-Poissonian statistics) as indicated by the crossing point. In Figs.~~\ref{fig_variance_anomalousXXZ}(b), (d), and (f), from the finite-size scaling analysis with $L=8,10,\cdots,18$, the anomalous scaling $V_\mathrm{anom}\propto L^\alpha$ is estimated for $\gamma=1$ as \ccred{$\alpha=2.64\pm0.02, 2.66\pm0.02, 2.67\pm0.02$} for XX and XXZ ($J_z=0.5, 1.5$) models. Remarkably, these values are close to each other including the exponent \ccred{$\alpha=2.67\pm0.02$} for the Heisenberg model and indicate the universality of SFQJ in many-body open quantum systems. We note that, in Fig.~\ref{fig_variance_anomalousXXZ}(b) for weak measurement strength, we find an oscillation of $V_\mathrm{anom}/L$ with respect to the system size, the behavior of which may stem from the specific integrability of the noninteracting XX chain in the unitary limit. We also remark that, as the average particle number thermalizes and is distributed homogeneously in long times, such an oscillation is not coming from the contribution of the initial N\'eel state.

\section{Numerical results for the analysis on the $\gamma$-dependence of $V_\mathrm{anom}$}
\begin{figure}[h]
\includegraphics[width=18cm]{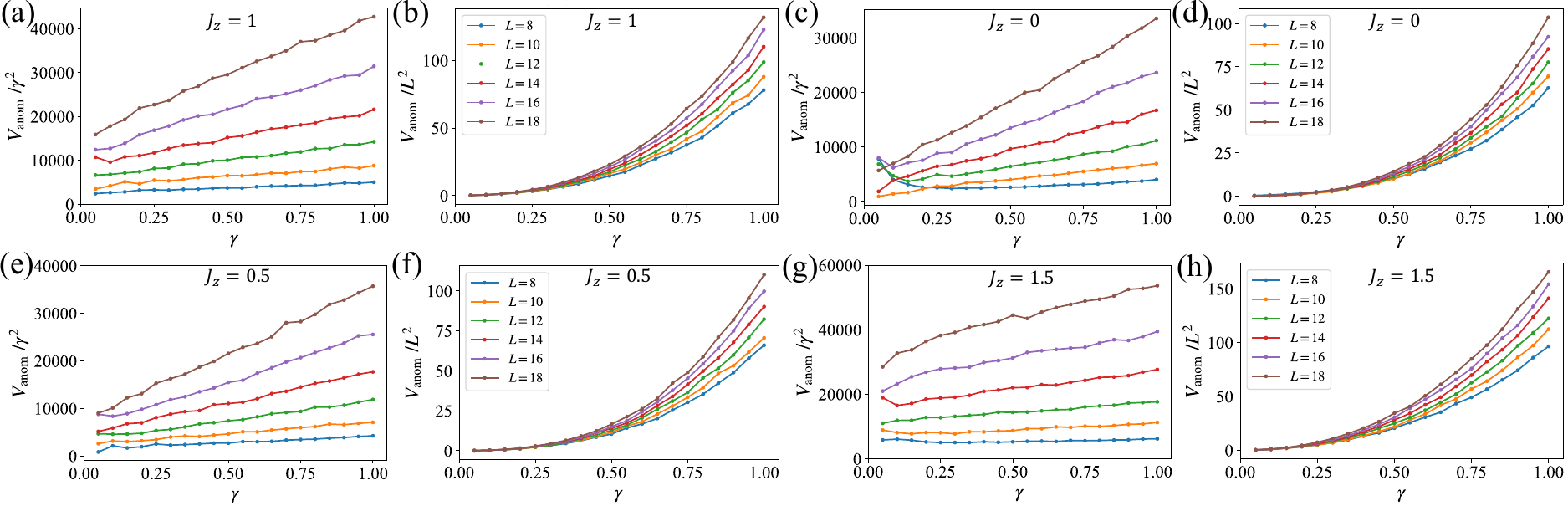}
\caption{Numerical results for $V_\mathrm{anom}$ in the half chain to analyze the $\gamma$- and $L$-dependence in (a), (b) Heisenberg, (c), (d) XX, (e), (f) XXZ ($J_z=0.5$), and (g), (h) XXZ ($J_z=1.5$) models for $832$ trajectories and $T=990$. (a), (c), (e), (g) $V_\mathrm{anom}/\gamma^2$ and (b), (d), (f), (h) $V_\mathrm{anom}/L^2$ are plotted against $\gamma$ for system sizes $L=8, 10, \cdots, 18$. We take the average over 20 time intervals for $t\in [200, 1190], [1190, 2180], \cdots, [19010, 2\times10^4]$.}
\label{fig_varianceanomgamma}
\end{figure}
We give additional numerical results to analyze the $\gamma$-dependence of $V_\mathrm{anom}$ in SFQJ for the half chain and also compare the behavior of $V_\mathrm{anom}$ with SFQJ. In Figs.~\ref{fig_varianceanomgamma}(a), (e), and (g), we find that $V_\mathrm{anom}/\gamma^2$ in the Heisenberg ($J_z=1$) and XXZ ($J_z=0.5,1.5$) model almost monotonically increases as the measurement strength is increased, but $V_\mathrm{anom}$ is not simply proportional to $\gamma^\beta$ with an exponent $\beta\:(>2)$ because $V_\mathrm{anom}/\gamma^2$ seems to become a positive constant as we approach the $\gamma\to0$ limit. On the other hand, in Fig.~\ref{fig_varianceanomgamma}(c) for the XX model ($J_z=0$), $V_\mathrm{anom}/\gamma^2$ seems to approach $0$ in the $\gamma\to0$ limit for odd $L/2$, while it shows the nonmonotonic behavior for even $L/2$ for small $\gamma$. We note that though the data for $L=18$ seem to deviate from $0$ in the $\gamma\to0$ limit due to numerical limitation, we still see that those for strong $\gamma$ are proportional to $\gamma$. Then, as the nonmonotonic behavior for even $L/2$ seems to vanish as the system size is increased, we predict that $V_\mathrm{anom}$ would become an increasing function of $\gamma$ for large system sizes possibly passing through the origin $(V_\mathrm{anom},\gamma)=(0,0)$. One possible scaling to explain the behavior in Figs.~\ref{fig_varianceanomgamma}(a), (c), (e), and (g) is $V_\mathrm{anom}\propto \gamma^2 c(\gamma)\propto\gamma^2 \sqrt{J_z^2+\gamma^2}$, but the data are rather rough to discuss the precise $\gamma$-dependence of $V_\mathrm{anom}$, and we need more sophisticated study. We also see in Figs.~\ref{fig_varianceanomgamma}(b), (d), (f), and (h) that the crossing found in $V[N_\mathrm{jump}^\mathrm{half}]/L^2$ [see Fig.~\ref{fig_variance}(c) in the main text, Figs.~\ref{fig_varianceXX}(c), \ref{fig_varianceXXZ0p5}(c), and \ref{fig_varianceXXZ1p5}(c)] does not appear in $V_\mathrm{anom}/L^2$. This means that the system-size scaling of $V_\mathrm{anom}$ does not change too much with respect to the measurement strength, in contrast with the measurement-induced crossover of SFQJ found in Fig.~\ref{fig_variance}(d) in the main text and in Figs.~\ref{fig_varianceXX}(d), \ref{fig_varianceXXZ0p5}(d), and \ref{fig_varianceXXZ1p5}(d). It is worth exploring the phenomena that exhibit a measurement-induced crossover or a measurement-induced phase transition in $V_\mathrm{anom}$ itself, but we leave it for future study.

\section{Further numerical results for the integrated autocorrelation functions and the Liouvillian gap}
\begin{figure}[h]
\includegraphics[width=17cm]{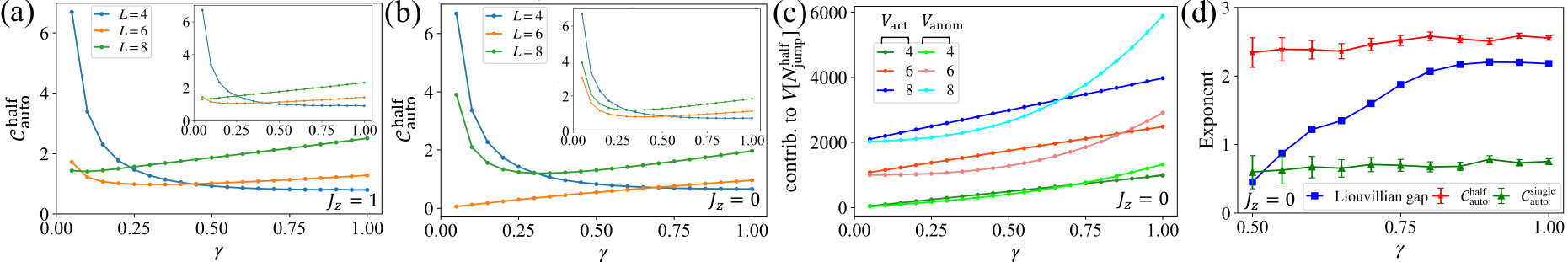}
\caption{Numerical results for the integrated autocorrelation functions and the Liouvillian gap for Heisenberg and XX models. (a) [(b)] Integrated autocorrelation functions for the half chain [Inset: $L/2\cdot\mathcal C_\mathrm{auto}^\mathrm{single}$] for the Heisenberg (XX) model and (c) $V_\mathrm{anom}$ and $V_\mathrm{act}$ for system sizes up to $L=8$ obtained with the use of the spectral decomposition \eqref{eq_variance_spc} in the main text. (d) Exponents $a$, $b$, and $c$ of the Liouvillian gap $\Delta\propto 1/L^a$ (blue, exact diagonalization), $\mathcal C_\mathrm{auto}^\mathrm{single}\propto L^b$ (green, quantum trajectory method), and $\mathcal C_\mathrm{auto}^\mathrm{half}\propto L^c$ (red, quantum trajectory method) obtained from the finite-size scaling analysis with $L=8, 10, 12, 14$. The leading contribution of $V_\mathrm{anom}$ and $V_\mathrm{act}$ changes at a critical measurement strength in (c) consistent with Fig.~\ref{fig_variance_anomalousXXZ}(a). The anomalous scaling of SFQJ in (d) cannot be captured by the Liouvillian gap. Data are shifted by $1000$ as $L$ is increased in (c) to improve visualization.}
\label{fig_autoexponentXX}
\end{figure}
We numerically investigate the integrated autocorrelation functions and the Liouvillian gap in the XX model and also give a supplemental figure for the Heisenberg model. First, we depict in Fig.~\ref{fig_autoexponentXX}(a) the integrated half-chain autocorrelation function for the Heisenberg model [Inset: $L/2\cdot\mathcal C_\mathrm{auto}^\mathrm{single}$] calculated from Eq.~\eqref{eq_variance_spc} in the main text. Though it seems that $\mathcal C_\mathrm{auto}^\mathrm{half}\sim L/2\cdot \mathcal C_\mathrm{auto}^\mathrm{single}$ is satisfied for the Heisenberg model up to $L=8$, this is a finite-size effect because $\mathcal C_\mathrm{auto}^\mathrm{half}\gg L/2\cdot \mathcal C_\mathrm{auto}^\mathrm{single}$ holds for larger system sizes. Moreover, we have numerically checked that off-diagonal terms $\mathcal C_\mathrm{auto}^{ij}\equiv\int_0^T d\tau\langle n_i^\prime(\tau)n_j^\prime\rangle_\infty$ ($i\neq j$) take positive- and negative-mixed values. Numerically, up to the system size $L=8$, off-diagonal terms $\mathcal C_\mathrm{auto}^{ij}$ cancel out with each other so as not to contribute to $\mathcal C_\mathrm{auto}^\mathrm{half}$. However, such terms give a non-negligible contribution for larger system sizes as shown in Fig.~\ref{fig_autocorrelation} in the main text. We also see that integrated autocorrelation functions become increasing functions of $\gamma$ for system sizes with $L\geq 8$. 

In Fig.~\ref{fig_autoexponentXX}(b), we show the integrated half-chain autocorrelation function for the XX model up to the system size $L=8$ obtained by using the spectral decomposition \eqref{eq_variance_spc} in the main text. In the inset, we depict $L/2\cdot\mathcal C_\mathrm{auto}^\mathrm{single}$ for comparison. We find that $\mathcal C_\mathrm{auto}^\mathrm{half}$ oscillates as the system size is increased for small $\gamma$ and $\mathcal C_\mathrm{auto}^\mathrm{half}\neq L/2\cdot \mathcal C_\mathrm{auto}^\mathrm{single}$ for $L=6$. This behavior may reflect the specific integrability of the free XX spin chain in the unitary limit. Moreover, we have numerically checked that off-diagonal terms $\mathcal C_\mathrm{auto}^{ij}$ ($i\neq j$) take positive- and negative-mixed values and can contribute to $\mathcal C_\mathrm{auto}^\mathrm{half}$. It is worth noting that, in Fig.~\ref{fig_autoexponentXX}(b) for odd $L/2$, $\mathcal C_\mathrm{auto}^\mathrm{half}$ goes to zero for $\gamma\to0$, which comes from the negative contribution of off-diagonal elements $\mathcal C_\mathrm{auto}^{ij}$. We also obtain $V_\mathrm{anom}$ and $V_\mathrm{act}$ with the use of Eq.~\eqref{eq_variance_spc} in the main text as shown in Fig.~\ref{fig_autoexponentXX}(c). We see that there exists a crossing of $V_\mathrm{anom}$ and $V_\mathrm{act}$ and the critical point $\gamma_c$ quantitatively agrees well with that in Fig.~\ref{fig_variance_anomalousXXZ}(a) for $L=8$.

Next, we compare the exponent of the Liouvillian gap with that of $\mathcal C_\mathrm{auto}^\mathrm{single}$. We remind that the asymptotic behavior of the local autocorrelation function is dominated by the Liouvillian gap as $|C_\mathrm{auto}^\mathrm{single}(\tau)|\sim e^{-\Delta \tau}$ $(\tau\to\infty)$ \cite{Mori23, Shirai24}, and if the scaling of the integrated autocorrelation function were solely determined by the Liouvillian gap, we could estimate it as $\mathcal C_\mathrm{auto}^\mathrm{single}\sim \int_0^T e^{-\Delta \tau}\sim 1/\Delta$. In Fig.~\ref{fig_autoexponentXX}(d), we depict the exponent of the inverse Liouvillian gap (blue curve) obtained from the exact diagonalization of the Liouvillian $\mathcal L$. We also show $\mathcal C_\mathrm{auto}^\mathrm{single}$ (green curve) calculated from the quantum trajectory method. By employing the finite-size scaling analysis with $L=8, 10, 12, 14$, we find that the Liouvillian gap is scaled as $\Delta\propto1/L^{2.18}$ for $\gamma=1$, which is close to the value $\Delta\propto1/L^2$ for $\gamma\gg1/L$ \cite{Cai13, Znidaric15}. On the other hand, for $\mathcal C_\mathrm{auto}^\mathrm{single}$, we obtain \ccred{$\mathcal C_\mathrm{auto}^\mathrm{single}\propto L^{0.75\pm0.04}$} for $\gamma=1$. This means that the scaling for the integrated autocorrelation function is not solely governed by the asymptotic dynamics described by the Liouvillian gap, but the transient dynamics before it significantly affects the scaling \cite{Shirai24}. As for the exponent of the integrated half-chain autocorrelation functions (red curve) obtained from the finite-size scaling analysis with $L=8, 10, 12, 14$, we find \ccred{$\mathcal C_\mathrm{auto}^\mathrm{half}\propto L^{2.55\pm0.03}$} for $\gamma=1$. We have numerically checked that the off-diagonal elements $C_\mathrm{auto}^{ij}$ with $i\neq j$ as well as diagonal components in $\mathcal C_\mathrm{auto}^\mathrm{half}$ give a finite contribution to the scaling as $\mathcal C_\mathrm{auto}^\mathrm{half}\gg L/2\cdot \mathcal C_\mathrm{auto}^\mathrm{single}$ for large $\gamma$. Thus, we cannot estimate the anomalous scaling exponent for $\mathcal C_\mathrm{auto}^\mathrm{half}$ given in Eq.~\eqref{eq_scaling} in the main text by the Liouvillian gap in the XX model, either.

\end{document}